# New Approach to Optimize the Time of Association Rules Extraction

Thabet Slimani
CS Department, Taif University, P.O.Box 888, 21974, KSA

**Abstract**
The knowledge discovery algorithms have become ineffective at the abundance of data and the need for fast algorithms or optimizing methods is required. To address this limitation, the objective of this work is to adapt a new method for optimizing the time of association rules extractions from large databases. Indeed, given a relational database (one relation) represented as a set of tuples, also called set of attributes, we transform the original database as a binary table (Bitmap table) containing binary numbers. Then, we use this Bitmap table to construct a data structure called Peano Tree stored as a binary file on which we apply a new algorithm called BF-ARM (extension of the well known Apriori algorithm). Since the database is loaded into a binary file, our proposed algorithm will traverse this file, and the processes of association rules extractions will be based on the file stored on disk. The BF-ARM algorithm is implemented and compared with Apriori, Apriori+ and RS-Rules+ algorithms. The evaluation process is based on three benchmarks (Mushroom, Car Evaluation and Adult). Our preliminary experimental results showed that our algorithm produces association rules with a minimum time compared to other algorithms.

## 1. Introduction

As a prominent tool for knowledge mining, Data mining [1] includes several techniques: Clustering, Association, Classification and Deviation. Knowledge Discovery in Data (KDD) constitutes an important advance in the area of data mining. It consists in the extraction of implicit knowledge (previously unknown and potentially useful), hidden in large databases. Association rule mining [2] is one of the principal problems treated in KDD and can be defined as extracting the interesting correlation and relation among huge amount of transactions. The task of association rule mining is to find interesting relationships from the data in the form of rules. The original application of association rule mining was on market basket analysis with the aim to study the buying habits of customers [3]. Currently, ARM has been the subject of several real-world applications in different areas requiring research groups of potential product or service, such as: medical diagnosis [4], biological database [5][6], electronic commerce [7][8] misuse detection [9].

Formally, an association rule is an implication relation in the form X→Y between two disjunctive sets of items X and Y. A typical example of an association rule on "market basket data" is that "80% of customers who purchase bread also purchase butter ". Each rule has two quality measurements, support and confidence. The rule X→Y has confidence c if c% of transactions in the set of transactions D that contains X also contains Y. The rule has a support S in the transaction set D if S% of transactions in D contain X∪Y. The problem of mining association rules is to find all association rules that have a support and a confidence exceeding the user-specified threshold of minimum support (called MinSup) and threshold of minimum confidence (called MinConf ) respectively.

Actually, frequent itemset mining and association rule mining became a wide research area in the field of data mining, and consequently a large number of quick and speed algorithms have been developed. The more efficient are those Apriori based algorithms or Apriori variations. The works that used Apriori as a basic search strategy, they also adapted the complete set of procedures and data structures [3][10][11]. Additionally, the scheme of this important algorithm was also used in sequential pattern mining [12], episode mining, functional dependency discovery & other data mining fields (hierarchical association rules [13]).

Another work was concentrated to develop faster algorithms for existing classical methods and adapting the algorithms into various states. As examples: multidimensional database mining [14], ontology based rule mining [15], association rule mining from the data cube [16][17][18], association rule mining in data warehouses [19], ontology based rule mining [20], parallel algorithms for association rule mining [21][22] and other algorithms.

Finding association rules is valuable for crossing-marketing [23] and attached mailing applications. Other applications include catalog design, add-on sales, store layout, and customer segmentation based on buying patterns. Besides application in the business area, mining association rule can also be applied to other areas, such as medical diagnosis [24], and remotely sensed imagery [25].

The databases involved in these applications are very large. Mining association rules in such databases may require substantial processing power. Therefore, it is necessary to have fast algorithms for this task. This observation motivates us to propose a new method for mining association rules in large databases. Given a relational database with various types of attributes (binary or not binary attributes), we first propose to convert the original database in a binary table (Bitmap). This transformation is a characteristic of the rough set method described in [26]. Next, we use a structure of data, called Peano tree (Ptree) which provides a lossless and compressed representation of Bitmap. By using Ptrees, an association rule mining algorithm with fast support calculation and significant pruning techniques are possible. The present work illustrates that using efficient data structures (Ptree) and our B-ARM algorithm (Binary Association Rule Mining) can be interesting for extracting the frequent itemsets[1] which is a time consuming task, especially when databases are large.

The Ptrees are used, in this context, to extend the Anding operation [27] of Ptrees to the attributes of the database. Using Ptrees, we do not make the expensive task of database scan each time we need to calculate the itemsets supports because, as we have already mentioned, concerned database is charged in a binary file, and Therefore we use the proposed B-ARM algorithm to extract the frequent itemsets on the specified Data Base in minimal time compared to other works.

This paper is organized as follows: in Section2, we describe briefly some related works. Section 3 summarizes the Ptree structure. Section 4 presents the specification of database attributes. In section 5, we details how to derive association rules using Ptrees. In Section 6, we describe our BF-ARM algorithm of the association rule mining. Details on implementation and experimental results are discussed in Section 7. Finally, we conclude with a summary of our approach and extensions of this work.

## 2. Literature Review

Early studies examined efficient mining association rules from different point of views. Apriori [28] is certainly the basic algorithm; it is developed for rule mining in large transaction databases. A DHP (Direct Hashing and Pruning) is an extension of the Apriori algorithm using a hashing technique [29]. A more recent algorithm called FDM (Fast Distributed Mining of association rules) was proposed by Cheung et al.[30], it is characterized by the generation of a small number of candidate sets and by the reduction of the number of messages to be passed at mining association rules. PincerSearch[31] spreads Apriori algorithm to generate the frequent itemsets. Depth-project [32] uses a dynamic reordering in order to reduce the research space. Another work realized by [33] proceeds to the improvement of the quality of the association rules by rough set technique. At least, nearer of our work, on one hand, FP-growth algorithm that represents the basis of transactions in the form of a compressed tree called FP-tree [34] and on the other hand, the MFItemsets algorithm (Maximum Frequent Itemsets) that represents the database as a truth table with an output Boolean function and sends back a body of Boolean products corresponding to the maximum frequent itemsets associated with the given transactions [35]. More recently, The work proposed by Rajalakshmi et al. [36] which identify maximal frequent itemsets based on minimum effort. The following paragraphs give a more detailed explanation of the previous approaches for more clarification:

**Apriori:** Apriori proposed by [28] is the fundamental algorithm. It searches for frequent itemset browsing the lattice of itemsets in breadth. The database is scanned at each level of lattice. Additionally, Apriori uses a pruning technique based on the properties of the itemsets, which are: If an itemset is frequent, all its sub-sets are frequent and not need to be considered.

**DHP:** DHP algorithm (Direct Haching and Pruning) proposed by [29] is an extension of the Apriori algorithm, which use the hashing technique with the attempts to efficiently generate large itemsets and reduces the transaction database size. Any transaction that does not contain any frequent k-itemsets cannot contain any frequent (k+1)-itemsets and such a transaction may be marked or removed.

**FDM:** FDM (Fast Distributed Mining of association rules) has been proposed by [30], which has the following distinct features.

1. The generation of candidate sets is in the same spirit of Apriori. However, some relationships between locally large sets and globally large ones are explored to generate a smaller set of candidate sets at each iteration and thus reduce the number of messages to be passed.
2. The second step uses two pruning techniques, local pruning and global pruning to prune away some candidate sets at each individual sites.
3. In order to determine whether a candidate set is large, this algorithm requires only O(n) messages for support count exchange, where n is the number of sites in the network. This is much less than a straight adaptation of Apriori, which requires $O(n^2)$ messages.

---
[1] Itemsets which have support above the user-specified minimum support.

**PincerSearch:** The Pincer-search algorithm [31] proposes a new approach for mining maximal frequent itemset which combines both bottom-up and top-down searches to identify frequent itemsets effectively. It classifies the data source into three classes as frequent, infrequent, and unclassified data. Bottom-up approach is the same as Apriori. Top-down search uses a new set called Maximum-Frequent-Candidate-Set (MFCS). It also uses another set called the Maximum Frequent Set (MFS) which contains all the maximal frequent itemsets identified during the process. Any itemset that is classified as infrequent in bottom-up approach is used to update MFCS. Any itemset that is classified as frequent in the top-down approach is used to reduce the number of candidates in the bottom–up approach. When the process terminates, both MFCS and MFS are equal. This algorithm involves more data source scans in the case of sparse data sources.

**Depth-project**: DepthProject proposed by Agarwal et al., (2000) [32] also mines only maximal frequent itemsets. It performs a mixed depth-first and breadth-first traversal of the itemset lattice. In the algorithm, both subset infrequency pruning and superset frequency pruning are used. The database is represented as a bitmap. Each row in the bitmap is a bitvector corresponding to a transaction and each column corresponds to an item. The number of rows is equal to the number of transactions, and the number of columns is equal to the number of items. By using the carefully designed counting methods, the algorithm significantly reduces the cost for finding the support counts.

**FP-tree :** FP-tree proposed by Han et al., (2000) [34] is a compact data structure that represents the data set in tree form. Each transaction is read and then mapped onto a path in the FP-tree. This is done until all transactions have been read. Different transactions that have common subsets allow the tree to remain compact because their paths overlap. the size of the FP-tree will be only a single branch of nodes. The worst case scenario occurs when every transaction has a unique itemset and so the space needed to store the tree is greater than the space used to store the original data set because the FP-tree requires additional space to store pointers between nodes and also the counters for each item.

**GenMax:** GenMax proposed by Gouda and Zaki, [37] a backtrack search based algorithm for mining maximal frequent itemsets. GenMax uses a number of optimizations to prune the search space. It uses a novel technique called progressive focusing to perform maximality checking, and diffset propagation to perform fast frequency computation.

**FPMax**: FPMax (Frequent Maximal Item Set) is an algorithm proposed by Grahne and Zhu, (2005) [38] based on FP Tree. It receives a set of transactional data items from relational data model, two interesting measures Min Support, Min Confidence and then generates Frequent Item Sets with the help of FPTree. During the process of generating Frequent Item Sets, it uses array based structure than tree structure. Additionally, the FPMax is a variation of the FP-growth method, for mining maximal frequent item sets. Since FPMax is a depth-first algorithm, a frequent item set can be a subset only of an already discovered MFI.

**Method based on minimum effort**: The work proposed by Rajalakshmi et al. (2011) [36] describes a novel method to generate the maximal frequent itemsets with minimum effort. Instead of generating candidates for determining maximal frequent itemsets as done in other methods [31], this method uses the concept of partitioning the data source into segments and then mining the segments for maximal frequent itemsets. Additionally, it reduces the number of scans over the transactional data source to only two. Moreover, the time spent for candidate generation is eliminated. This algorithm involves the following steps to determine the MFS from a data source:
1. Segmentation of the transactional data source.
2. Prioritization of the segments
3. Mining of segments

## 3. Ptree Structure

The data structure tree Peano (Ptree), also called "peano count tree" is a compact and efficient representation used to store a database (originally an image) as a binary bits (0 and 1). This structure was initially introduced for the representation of spatial data such as RSI data applications (Remotely Sensed imagery) [39][27]. Using Ptree structure, all the count information can be calculated quickly. This facilitates efficient ways for data mining.

A Ptree is a quadrant based tree. The Ptree principle is to divide, recursively, the totality of spatial data into quadrants and counting the bits having the value "1" for each quadrant, thus forming a computation quadrants tree. In figure 1, 55 is the number of bits in one complete picture, the root level is labeled level 0. The numbers at the next levels (level 1) are, 16, 8, 15 and 16, are the 1-bit counts for the four major quadrants. The quadrants composed entirely of 1-bits are called a "pure 1 quadrant" (the first and last quadrant with 16 value are a pure 1 quadrant and we do not need sub-trees for these two quadrants, so these branches terminate) and similarly, the quadrants composed entirely of 0-bits are called a "pure 0 quadrant" (which also terminate). This process is repeated recursively using the Z-ordering of the four sub-quadrants at each new level. Eventually, every branch terminates in the leaf level or when each quadrant is a pure quadrant.

The Ptrees are similar in their construction to other existing data structures, for example Quadtrees (Samet 1984) and HHcodes[1]. The similarities between Ptrees, quadtrees, and HHcodes[2] are that they are quadrant based. The difference is that Ptree include occurrence counts. Trees are not indexed, but they are representations of the dataset itself.

When using the Ptree structure, any information calculation can be completed very fast. The performance analysis realized in [27] shows that Ptree produces a good cost computation (CPU time) and reduce the storage space compared to the original data.

Peano mask tree (pm-tree) is a variation of the ptree data structure. Pm-tree is a similar structure in which masks rather than counts are used. In pm-tree structure, to represent pure-1, pure-0 and mixed quadrant we use a 3-value logic. In pm-tree structure, to represent pure-1, pure-0 and mixed quadrant we use a 3-value logic. Pm-tree is helpful for the optimization of anding operation between two ptrees.

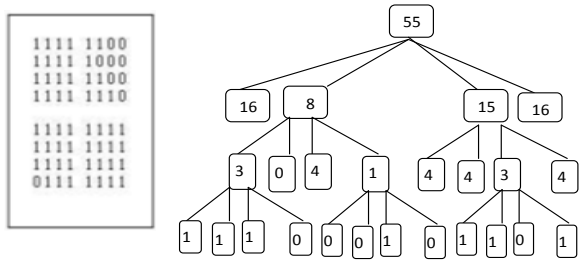

Figure1. Ptree for 8*8 image

A PM-tree example is given in the Figure 2. Other variations can be used, such as P1-tree and P0-Tree. In P1-tree, we use 1 to indicate the pure-1 quadrant while use 0 to indicate others. In P0-tree, we use 1 to indicate the pure-0 quadrant while use 0 to indicate others. Both P1-tree and P0-tree are lossless representations of the original data [39].

---

[1] http ://www.statkart.no/nlhdb/iveher/hhtext.html
[2] http ://www.statkart.no/nlhdb/iveher/hhtext.html

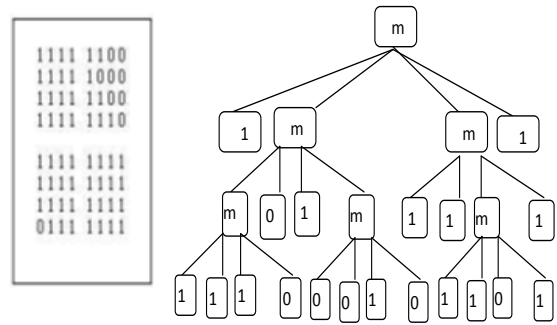

Figure2. PMtree for 8*8 image

## 4. Specification of Database Attributes

A Database is represented by a binary or bitmap table whose columns are attributes and each attribute owns a limited set of values (items) known by the domain attributes of the database. A database can have two types of attributes domain: Binary attributes domain (BAD) and non-binary attributes domain (NBAD).

**Binary Attributes domain**: A Binary attributes domain is represented by a vector $\vec{v} \subset \{v1, v2\}$ with size k, such that the values v1 and v2 are taken from the set {0,1}, and k is the number of k-tuples of values taken from {0, 1}. 1-tuple represents a tuple of the database or transaction in terms of the market basket data analysis.

A database $A^n$ n-dimensional is constituted of n binary vectors, when each vector has $2^n$ size and is constituted in turn with 4 binary vectors $2^n/4$ (for simplicity reasons, we decompose each binary vector into four quadrants).

$A^n$ Lines (transactions) repesent all combinations of n possible binary values 0 and 1. In the example given in table.1, the presence
of a computer item in a transaction or its absence represents its domain {purchased, not purchased} and the binary transformation makes the attribute value a1 =1, if the computer is purchased or a1 = 0 if the computer is not purchased.

**Table1.** The transformation of raw data into a bitmap representation for BAD.

| Tid | Computer | | Tid | a1 |
|---|---|---|---|---|
| 1 | Purchased | ➔ | 1 | 1 |
| 2 | Not purchased | | 2 | 0 |
| 3 | Purchased | | 3 | 1 |
| 4 | Not purchased | | 4 | 0 |
| …. | …. | | ….. | ….. |

**Non Binary Attributes domain**: A non Binary attribute domain Aj is consituted with j items of the Database and represented by $\sum_{i=1}^{n} j*i$ binary vectors where n is the number of attributes of the non binary attributes domain. For example, for a better representation of the benefit of a client, we associated to the attribute "income" the domain with three (j=3) items {high, medium, low} defined as follows: a1 = "high income" a2 = "middle income" and a3 = "low income" and represented by the following binary table (Table 2):

**Table 2.** The transformation of raw data into a bitmap representation for NBAD

| Tid | Income |   | Tid | a1 | a2 | a3 |
|---|---|---|---|---|---|---|
| 1 | High | → | 1 | 1 | 0 | 0 |
| 2 | Meduim |   | 2 | 0 | 1 | 0 |
| 3 | Low |   | 3 | 0 | 0 | 1 |
| 4 | High |   | 4 | 1 | 0 | 0 |
| …. | ……….. |   | ….. | … | … | … |

## 5. Association Rule Mining using Ptree

Given a user-specified minimum support and minimum confidence, the problem of mining association rules is to find all the association rules whose support and confidence are larger than the respective thresholds specified. Thus, it can be decomposed into two subproblems :
Finding the frequent itemsets which have support above the user-specified minimum support.
Deriving all rules, based on each frequent itemset, which have more than user-specified minimum confidence.
The whole performance is mainly determined by the first step, which is the generation of frequent itemsets. Once the frequent itemsets have been generated, it's straightforward to derive the rules. To solve this problem and to improve the performance, Apriori algorithm was proposed [28]. Apriori algorithm generates the candidate itemsets to be counted in the pass by using only the itemsets found large in the previous pass - without considering the transactions in the database.
The key idea of Apriori algorithm lies in the "downward-closed" property of support which means if an itemset has minimum support, then all its subsets also have minimum support. An itemset having minimum support is called frequent itemset (also called large itemset). So any subset of a frequent itemset must also be frequent. The candidate itemsets having k items can be generated by joining frequent itemsets having k-1 items, and deleting those that contain any subset that is not frequent.

Start by finding all frequent 1-itemsets (itemsets with 1 item); then consider 2-itemsets, and so forth. During each iteration only candidates found to be frequent in the previous iteration are used to generate a new candidate set during the next iteration. The algorithm terminates when there are no frequent k-itemsets. Since finding frequent itemsets is a time consuming task, especially when databases are large, we use a Bitmap table (containing binary data) to organize original database and the Peano tree (Ptree) structure to represent Bitmap tables in a spatial data mining-ready-way. Ptrees are a lossless representation of the original database.

### 5.1. Model Representation

The rough set method [26] operates on data matrices, called "information tables" which contain data about the universe $\Omega$ of interest, condition attributes $\Omega c$ and decision attributes $\Omega d$. The goal is to derive rules that give information how the decision attributes depend on the condition attributes.
Let us consider the original database represented by the corresponding information contained in Table 3 where condition attributes are {A, B, C, D, E, F} associated to the following products {cartridge printer, video reader, car, computer, movie camera, printer} and decision attribute is {G} corresponding to {graphic software}. Each attribute has two non null values {yes, no}. So, there are seven items (bitmap-attributes) for the resulting bitmap-table {A, B, C, D, E, F and G}.

**Table 3. Original table (database) with its equivalent Bitmap table.**

| $T_{id}$ | A | B | C | D | E | F | G |
|---|---|---|---|---|---|---|---|
| $T_{id1}$ | yes | yes | no | no | no | no | no |
| $T_{id2}$ | yes | yes | yes | yes | yes | yes | no |
| $T_{id3}$ | no | yes | no | yes | no | no | yes |
| $T_{id4}$ | no | yes | no | no | yes | no | yes |
| $T_{id5}$ | no | no | no | yes | no | yes | yes |
| $T_{id6}$ | no | no | no | yes | yes | no | yes |
| $T_{id7}$ | no | yes | no | no | yes | no | no |
| $T_{id8}$ | no | yes | no | yes | yes | yes | no |

| Tid | A | B | C | D | E | F | G |
|---|---|---|---|---|---|---|---|
| $T_{id1}$ | 1 | 1 | 0 | 0 | 0 | 0 | 0 |
| $T_{id2}$ | 1 | 1 | 1 | 1 | 1 | 1 | 0 |
| $T_{id3}$ | 0 | 1 | 0 | 1 | 0 | 0 | 1 |
| $T_{id4}$ | 0 | 1 | 0 | 0 | 1 | 0 | 1 |
| $T_{id5}$ | 0 | 0 | 0 | 1 | 0 | 1 | 1 |
| $T_{id6}$ | 0 | 0 | 0 | 1 | 1 | 0 | 1 |
| $T_{id7}$ | 0 | 1 | 0 | 0 | 1 | 0 | 0 |
| $T_{id8}$ | 0 | 1 | 0 | 1 | 1 | 1 | 0 |

In the first step, all possible rules are constructed from all bitmap-attributes of the table. All rules not fulfilling the minimum support (MinSup=10%) and minimum confidence (MinConf=30%) should be deleted.

## 5.2. From Attributes to Ptree

According Ptree representation, the set of tuples must be a power of 2. The number of the tuples in a database is transformed to the nearest power of 2, knowing than the completed tuples are itemsets uniquely including the value 0. In Ptree approach, each column of Bitmap table is represented by a vector of bits of which cuts it divisible by 4, called basic Ptree.

For simplicity reasons, we suppose that the fan-out is four. For every vector of bits, a basic Ptree is associated. There are six basic Ptrees for the universe $\Omega_c$ of Table.2, and since each Ptree presents a number of bits divisible by 4, therefore, it is constituted by four under quadrants of which the origin quadrant is the entirety of the bits forming Bitmap table. As the header of Ptree files contains the root count, the root counts of Ptrees are immediately accessible, and will conveniently replace the necessity of using original data to count the number of transactions containing candidate frequent itemsets.

must have at minimum 4 tuples). For example, if the number of tuples is less than 16, one completes by 0 to obtain the Ptree format on 16 tuples. Generally, if the number of transactions is inferior to $2^{n+1}$ and superior to $2^n$ then the basic Ptree is stored with $2^{n+1}$ number.

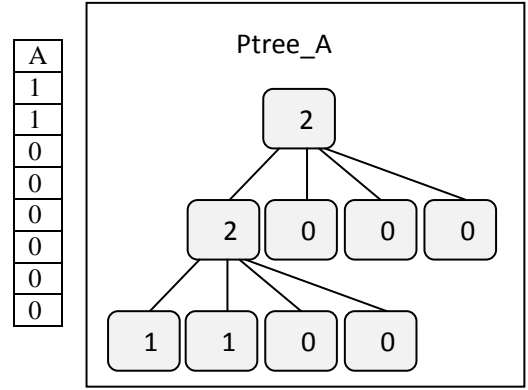

**Figure 4. Snaphsot of Ptrees representation (Ptree_A).**

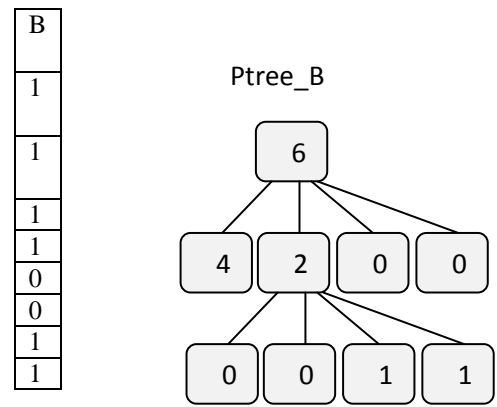

**Figure 5. Snaphsot of Ptrees representation (Ptree_B).**

| Tid | A | B | C | D | E | F | G |
|---|---|---|---|---|---|---|---|
| $T_{id1}$ | 1 | 1 | 0 | 0 | 0 | 0 | 0 |
| $T_{id2}$ | 1 | 1 | 1 | 1 | 1 | 1 | 0 |
| $T_{id3}$ | 0 | 1 | 0 | 1 | 0 | 0 | 1 |
| $T_{id4}$ | 0 | 1 | 0 | 0 | 1 | 0 | 1 |
| $T_{id5}$ | 0 | 0 | 0 | 1 | 0 | 1 | 1 |
| $T_{id6}$ | 0 | 0 | 0 | 1 | 1 | 0 | 1 |
| $T_{id7}$ | 0 | 1 | 0 | 0 | 1 | 0 | 0 |
| $T_{id8}$ | 0 | 1 | 0 | 1 | 1 | 1 | 0 |

**Figure 3. Tabular Representation of Ptrees.**

In a similar way to the generation of the sequence of Peano from the spatial data, we create the Ptree in a Bottom-up way. The generation of Ptree depends on the number of fan-out in the internal nodes of the Ptree and in the root node. To represent Ptree with different fans-outs, we introduce the Ptree-(r-i) notation ; where r = the fan-out of the root node and i = the fan-out of all the internal nodes of the level 1. We adopt in our work, the representation Ptree-(4-4-n), it means that this structure divides the database tuples into 4 blocks (the block of a transaction

## 5.3. Ptree Anding Operation

ANDing is a very important and frequently used operation for Ptrees. There are several ways to execute Ptree ANDing operation. We can execute anding level-by-level starting from the root level. Table. 4 gives the rules for performing Ptree ANDing. Operand 1 and Operand 2 are two ptrees with root $X_1$ and $X_2$ respectively. Using PM-trees, $X_1$ and $X_2$ could be any value between 1, 0 and m (3-value logic representing pure-1, pure-0 and mixed quadrant). For example, a pure-1 Ptree combined with any ptree will have as consequence the second operand and a pure-0 ptree with any ptree will result in the pure-0 ptree.

Table 4. Ptree anding rules.

| Operand 1 | Operand 2 | Result |
|---|---|---|
| 1 | $X_2$ | Subtree with root $X_2$ |
| 0 | $X_2$ | 0 |
| $X_1$ | 1 | Subtree with root $X_1$ |
| $X_1$ | 0 | 0 |
| m | m | 0 if four sub-quadrants result in 0 value; m otherwise |

## 6. Algorithm BF-ARM (Binary File Association Rule Mining)

In accordance with the Apriori algorithm, we wish to find all frequent 1-itemsets first. We do not need to scan the entire data set and employ a counter for each item. We only need to access the root count of the Ptree for every item in the table. Then a simple calculation will indicate whether the item is a frequent 1-itemset or not. If the root count of an item's Ptree divided by the total number of transactions is greater than Minsup, then the item is a frequent 1-itemset.

Using Ptrees structure has now saved one scan of the entire data set, along with the necessity of memory buffer management. Next we want to find frequent 2-itemsets. The only possible candidate 2-itemsets are made up of frequent 1-itemsets. All other Ptrees will be ignored in finding frequent 2-itemsets, so we are working with a subset of the original data set. The candidate 2-itemsets are all the pairwise combinations of frequent 1-itemsets. For each candidate 2-itemset, the Ptrees for the two items are ANDed to produce a third, derived Ptree that represents the presence of both items in transactions. Note that the ANDing process is fast enough that there is no need to save the new Ptree for further operations (Ding et al, 2002). To determine support for any candidate 2-itemset, we divide the root count of the new, 2-itemset Ptree by the number of transactions in the table. If support is greater than Minsup, we'll calculate the confidence levels to test against Minsup. The numbers needed to calculate confidence are, once again, the root counts of Ptrees.

Now, knowing the frequent 2-itemsets we can continue with discovering frequent 3-itemsets from candidate 3-itemsets and so on until no candidate k-itemsets exist. The basic algorithm is to AND all the Ptrees of the items in the candidate k-itemset, divide the root count of the new Ptree by the total number of transactions, and test for Minsup. In every case, there is no need to scan the data since the necessary counts exist already within the Ptrees derived from AND operations.

**Algorithm** *BF-ARM*
**Data Discretization**
**Ptrees_Storage**
**For each** attribute $i \subset \Omega_c$
  $C_1 = F_1$
**End For**
$C_k = C_1$
**Do While** ($C_k \neq \emptyset$)
  **For each** attribute $i \subset C_k$
    **For each** attribute $j \subset \Omega_d$
      $F_{ij}$ = AND_Ptreebase(i,j)
      **Storage_Ptrees**
    **End For**
    $F_k = F_k \cup F_i$  //itemsets candidats
  **End For**
  $C_k = F_k \{c \in C_k | c.count \geq MinSup\}$
**End While**

**Function Storage_Ptrees**
  **For** (bandj=1; j<I; j++)
    **root**[j] :=rootcount(1 ; bandj)
  //vector storing the roots of Ptrees.
    **If**($2^n \leq N_t$ and $N_t < 2^{n+1}$) **then**
      **For** (i :=$N_t$ ; i$\leq 2^{n+1}$ ; i++)
        bandj[i+1] :=0 ;
      **End For**
    **End If**
    k :=0 ;
  **For** (i :=1 ; i$\leq 2^n$; $2^n/4$)
    k :=k+1 ;
    **rootsBandj**[k] :=rootcount (1 ; bandj) ;
  **End For**
  **For** (i :=1 ; i$\leq 2^n$; $2^n/4$)
    **If** (rootsBandj[i]<>$2^n$ or rootsBandj[i]<>0)
     **then**
      **bitsBandj**[i] :=rootsBandj[i] ;
    // bits vectors
    **End If**
  **End For**
**End For**

The *rootcount* function, used in the storage procedure of Ptree, is to calculate admissible itemsets counts directly by ANDing the appropriate basic Ptrees instead of scanning the original databases. Let $N_t$ be the number of tuples; n is initialized to 3 and I be the total number of attributes.

The list of the strong rules generated in the example described previously is summarized in Table 5. The rules are classified by decision attribute G or F.

**Table 5. Strong rules generated by the BF-ARM algorithm.**

|  | Decision Attribute = G | Decision Attribute = F |
|---|---|---|
| Strong rules with 2 attributes | B→G<br>D→G<br>E→G<br>F→G | A→F<br>B→F<br>D→F<br>E→F |
| Strong rules with 3 attributes | B,D→G | A, B→F<br>A,C→F<br>A,D→F<br>A,E→F<br>B,C→F<br>B,D→F<br>C,D→F |
| Strong rules with 4 attributes |  | A,B,C→F<br>A,B,D→F<br>A,B,E→F<br>A,C,D→F<br>B,C,D→F<br>B,C,E→F<br>B,D,E→F<br>C,D,E→F |
| Strong rules with 5 attributes |  | A,B,C,D→F<br>A,B,C,E→F<br>A,C,D,E→F<br>B,C,D,E→F |
| Strong rules with 6 attributes |  | A, B,C,D,E→F |

## 7. Experimental Results

In this section, we compare our work with the two approaches Apr+ (Hybrid Association Rule Algorithm Apriori+) and Rs+ (Rough Set Based Rule Generation Algorithm RS-rules+) (Delic et al. 2002). The procedure Apr+ is an extension of the method "faster association rule" combined with the procedure "rough set". The derived rules are produced on the basis of the successive reduction of the useless rules. If there is a given fixed decision attribute, all itemsets without this attribute can be ignored for rule generation. Besides the Apr+ procedure of rule generation with a fixed decision attribute, the procedure Rs+ offers the possibility of varying the selected decision attributes, so, each attribute can be included either as a decision or condition attribute. In our work, we use the principle for deriving rules with a fixed decision attribute, but we add the notion of the Ptree structure to accelerate the processor time generation of the strong rules. Furthermore, in our work, we don't produce redundant rules, while in Rs+ and Apr+, redundant rules are produced and removed by the continuation.

The comparison of our work is facilitated by the use of a benchmark data set[1] concerning Car Evaluation Database, Mushroom Database and Adult database. For example, the car evaluation database contains 1728 tuples and 25 values of attributes (items) in the Bitmap table.

The preliminary experimental results given in Table 6 show that the computing times were in favor of our BF-ARM algorithm (Figure 6). Indeed, bases on Car Evaluation benchmark, our work produces a CPU time to generate strong rules equal to 0,083 min for fixed decision attributes and a CPU time equal to 0,067 min for no fixed decision attributes. By running Apr+, Rs+, and BF-ARM algorithms, we got identical rules.

**Table 6. Comparative table for the algorithms RS-ules+ (Rs+), Apriori+ (Apr+), Apriori (APR) and BF-ARM.**

| Database | Car Evaluation CPU Time(Min) | | Mushroom CPU Time(Min) | | Adult CPU Time(Min) | |
|---|---|---|---|---|---|---|
| MinSupp | 10% | | 35% | | 17% | |
| MinConf | 75% | | 90% | | 94% | |
| | Fixed Decision Attribute | | | | | |
| Method | Yes | No | Yes | No | Yes | No |
| **RS+** | 1.15 | 3.15 | 3.32 | 15 | 64 | 233 |
| **APR+** | 1.12 | 1.12 | 2.02 | 2.02 | 44 | 44 |
| **APR** |  | 1.10 |  | 2 |  | 44 |
| **BF-ARM** | 0.083 | 0.067 | 0.3 | 0.223 | 6.29 | 4.54 |

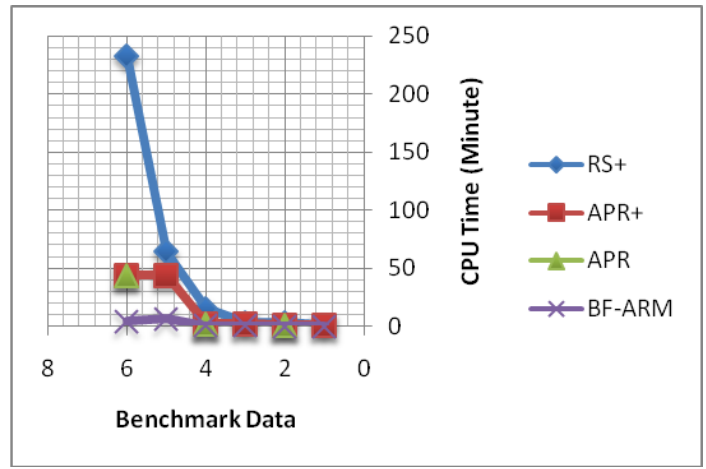

**Figure 6. Comparative Snapshot for the algorithms RS-rules+ (Rs+), Apriori+ (Apr+), Apriori (APR) and B-ARM.**

## 8. Conclusion

In this paper, we propose a new method to derive association rules from large databases using Ptree structure. The Ptree structure is a space efficient, lossless, data mining ready structure for binary datasets. The adopted association rules method specified by our algorithm BF-ARM is based on the concept of pruning by minimum support and minimum confidence directly to produce strong association rules. The discovery of similarities between attributes/items was based on the rules of Ptrees ANDing. Based on the benchmark data, we compared the quality of the produced rules and the necessary computing times of the algorithms. It turned out the generated rules of Apr+, Rs+, Apr and BF-ARM does not differ. But, the computing times were in favor of BF-ARM algorithm. Our work is beneficial because on the one hand, it avoids the direct scanning of database (an expensive operation in memory and computation time), which greatly exceeds the capacity of computers, despite their rapid evolutions and, secondly, it provides a gain of attributes comparison because the comparison is performed by a block of tuples. Another interesting direction is the extension of our association rule mining method by adding time constraints (new area for sequence identification).